# Precise H 1S-nS in Boltzmann-Hund double well potential and Sommerfeld-Dirac theories probe chiral behavior of atom H

G. Van Hooydonk, Ghent University, Faculty of Sciences, Ghent, Belgium

Abstract. *We consider generic Boltzmann-Hund double well potential (DWP) theory and find that, for atom H, the Sommerfeld-Dirac equation in QED transforms in a DWP (Mexican hat curve) for chiral systems. With either theory, the few precise H 1S-nS terms available probe the chiral nature of atom H.*
Pacs: 03.65.Pm

## I. Introduction

By comparing 1S-2S for H ($e^-p^+$) and for mirror $\underline{H}$ ($e^+p^-$) [1-2], theorists hope to understand CPT. However, Sommerfeld-Dirac equation (SDE) [3] gives fluctuations of ±14 kHz [4], too large to interpret H 1S-2S, precise to 47 Hz [5], while $\underline{H}$ remains problematic until H data are confronted with double well potential (DWP) theory for chiral systems [6]. We now prove that SDE and DWP theories are almost degenerate. With either theory for precise 1S-nS [5,7-11], we find a Mexican hat curve with H- and $\underline{H}$-states, which proves that atom H is chiral indeed [12-13].

## II. Theory

*II.1  Generic Boltzmann-Hund type DWP theory for chiral (quantum) systems*

Fig. 1 shows that dimensionless DWP in x or y=1-x describes a generic Mexican hat curve [6,14]

$$M_x = (1-x^2)^2 = 1-2x^2+x^4 = (2y-y^2)^2 = 4y^2-4y^3+y^4 = y^2(2-y)^2 = 4y^2(1-\tfrac{1}{2}y)^2 \qquad (1)$$

Achiral or (too) symmetric systems have one minimum. Chiral or less symmetric systems (1) have two, one for a left-, another for a right-handed system, with a maximum in between. As soon as simple $M_x$ (1) is detected in the simplest spectrum of simplest atom H, this is also prototypical for chiral behavior. This straightforward conclusion is tested below with precise H 1S-nS [5,7-11].

*II.2 Boltzmann-Hund type Mexican hat curve as contained in SDE for H bound state QED*

$M_x$ (1) is hidden in SDE, the cornerstone of bound state QED [3]. In standard notation [3], SDE

$$E(n,j) = \mu c^2[f(n,j)-1]; \quad f(n,j) = (1/\sqrt{\{1+\alpha^2/[\sqrt{((j+\tfrac{1}{2})^2-\alpha^2)}+n-(j+\tfrac{1}{2})]^2\}}) \qquad (2)$$

gives the spectrum of Dirac Hamiltonian for relativistic H. For nP, expansion to order $\alpha^4$ gives

$$-E_{nP} = \tfrac{1}{2}\mu\alpha^2 c^2(1/n^2+\alpha^2/n^3-\tfrac{3}{4}\alpha^2/n^4)+\text{h.o.} = (R_H/n^2)[1+\tfrac{1}{3}\alpha^2 - \tfrac{1}{3}\alpha^2(1-1{,}5/n)^2]+\text{h.o.} \qquad (3)$$

Since the parabolic part between square brackets is like (1), even SDE suggests chiral H behavior. In fact, (2) gives (3) using zero $0 = +\tfrac{1}{3}R_H\alpha^2/n^2 - \tfrac{1}{3}R_H\alpha^2/n^2$, pending h.o. Reduced difference

$$\Delta E_{nP}/(\tfrac{1}{3}R_H\alpha^2) = (-[E_{nP}+(R_H/n^2)(1+\tfrac{1}{3}\alpha^2)]/(\tfrac{1}{3}R_H\alpha^2) = (1/n^2)(1-1{,}5/n)^2 \qquad (4)$$

is the H Mexican hat curve, as contained in SDE: a plot of (4) versus (3/n-1) gives $M_x$ in Fig. 1. For $2S_{1/2}$, expected to be degenerate with $2P_{1/2}$, (2) was disproved [15]. We show in Appendix A that DWP and SDE are degenerate within parts in $10^{16}$ [5] and adapt (3) for $E_{nS}$ in Appendix B.

*II.3 Boltzmann-Hund chiral DWP $H_{nS}$ theory and H Mexican hat curve*

If achiral nP term in (3) is $E^a_{nP}=-R_H(1+1/3\alpha^2)/n^2$, $E^a_{nS}=-R_H(1+a)/n^2$ is expected for nS. With nP-symmetry broken by $E^c_{nP}=+1/3\alpha^4(1-n_{c(P)}/n)^2/n^2$, H nS-symmetry breaking will follow

$$E^c_{nS}=+aR_H(1-n_{c(S)}/n)^2/n^2 \qquad (5)$$

i.e. Mexican hat curve (1) [13], where a is of order $\alpha^4\mu c^2$ or $\alpha^2 R_H$. In DWP $H_{nS}$ theory

$$-E_{nS(DWP)}=E^a_n-E^c_n=(R_H/n^2)\{1+a[1-(1-n_c/n)^2]/n^2\}=R_H/n^2+2an_c/n^3-an^2_c/n^4 \qquad (6)$$

critical $n_{c(S)}$ depends solely on the ratio of terms in $1/n^3$ and $1/n^4$ in (5). With Lamb shifts [15], $n_{c(S)} \neq n_{c(P)}$. Left-right differences are small: achiral H behavior follows $\alpha^2$; chiral H behavior, its fine structure, follows $\alpha^4$. Fitting observed H terms

$$T_{nS}=\Delta E=|E_{nS}-E_{1S}| \qquad (7)$$

with Boltzmann-Hund chiral DWP theory therefore rests on $E_{nS}$ and $T_{nS}$ equations

$$-E_{nS(DWP)}=R_H/n^2+A/n^3-\tfrac{1}{2}An_c/n^4 \text{ cm}^{-1} \text{ and } T_{nS}=E_{1S}-R_H/n^2-A/n^3+\tfrac{1}{2}An_c/n^4 \text{ cm}^{-1} \qquad (8)$$

$$-nE_{nS(DWP)}=R_H/n+A/n^2-\tfrac{1}{2}An_c/n^3 \text{ cm}^{-1} \text{ and } nT_{nS}=nE_{1S}-R_H/n-A/n^2+\tfrac{1}{2}An_c/n^3 \text{ cm}^{-1} \qquad (9)$$

$$R_{nS}=-n^2E_{nS(DWP)}=R_H+A/n-\tfrac{1}{2}An_c/n^2 \text{ cm}^{-1} \text{ and } n^2T_{nS}=n^2E_{1S}-R_H-A/n+\tfrac{1}{2}An_c/n^2 \text{ cm}^{-1} \qquad (10)$$

wherein zero term in (3) is invisible. $R_{nS}$ (10) and the Mexican hat curve in (5) were found earlier [12,13] but not yet with precise $T_{nS}$. Which relation (8)-(10) to use rests on the number of precise $T_{nS}$ available: (8) needs 5 in 4th order $1/n$, (9) 4 in 3d order. With just 3 precise terms (see Section III), only (10) can give a mathematically exact fit for available data.

**III. Precise H 1S-nS intervals available**

Table 1 gives observed (I) and derived (II) intervals, uncertainties σ and differences d with recent [16] and older QED [17] (see [18]). For 3S and 6S, only difference D is known. Only relation (10) can give an exact fit for 2S, 4S and 8S. A prediction for 3S awaits its measurement [19]. SDE for precise 2S, 4S and 8S is in Appendix B, whereas QED for $H_{nS}$ is discussed in Appendix C.

**IV. Boltzmann-Hund (BH) chiral H theory and precise H nS levels**

$T_{2S}$, $T_{4S}$ and $T_{8S}$ with σ=0 can be fitted exactly with a parabola in $1/n$ but the result is not reliable for the complete nS-series. However, it gives a provisional $E_{1S}$ value

$$-E_{1S} \approx 29979245800.109678,744663\ldots \text{ Hz} \qquad (11)$$

the starting point for an extended analysis to order $1/n^4$. We distinguish between σ=0 and σ≠0.

*IV.1 Case 1: all σ=0*

(i) Using (11), DWP parabola (10) leads to running Rydberg differences

$$\pm\Delta R_{nS}=\pm[n^2(-E_{1S}-T_{nS})-R_H]\approx\pm[n^2(T_{nS}-109678,74)-109678,74]=\pm A(1/n-\tfrac{1}{2}/n^2) \text{ cm}^{-1} \qquad (12)$$

with a maximum or a minimum. Any $-E_{1S}=R_H$ value in (12) reproduces exactly the 3 input data but only one can also accurately reproduce term D for 3S and 6S, not used as input.

(ii) Calculating D for $-E_{1S}$, close to (11), gives the best $E_{1S}$-value (13) by interpolation. We get

$$-E_{1S}=R_H=109678{,}77174371161\ldots \text{cm}^{-1} = 3288086857146820 \text{ Hz} \qquad (13)$$

$$-\Delta R_{nS}=29979245800(4{,}368693379685/n^2-5{,}55578213674/n+1{,}188083721092) \text{ Hz} \qquad (14)$$

$$T_{nS}=29979245800[(109678{,}77174+3{,}71161.10^{-6})(1-1/n^2)+4{,}368693379685/n^4$$
$$-5{,}55578213674/n^3+1{,}188083721092/n^2] \text{ Hz} \qquad (15)$$

Parabola (14) in Fig. 2 leads to $T_{nS}$ (15), in exact agreement with observation, see Table 2. Fig. 2 shows Bohr $R_H=-E_{1S}$ (13), SDE $R_{\infty S}=109677{,}583659991$ cm$^{-1}$ but, at $n_{c(S)}=1{,}572665476$, also

$$R_{harm}=29979245800*109679{,}350018514 \text{ Hz} \qquad (16)$$

the harmonic Rydberg $R_{harm}$ [12]. These 3 H asymptotes lead to 3 possible Mexican hat curves, to be distinguished in Section VI. Reminding [21] for redirecting problems with $E_{1S}$, we see that

(a) although (13)-(15) use $\sigma=0$, they give $T_{1S}\neq 0$;

(b) BH fitting with QED value $-E_{1S}=109678{,}771743070$ cm$^{-1}$ [16] removes most QED errors in Table 1, leads to an (allowed) error of +9795 Hz for D but, just like (a), gives $T_{1S}\neq 0$ and

(c) using $T_{1S}=0$ gives $E_{1S}=109678{,}77174687022$ cm$^{-1}$, which returns input terms exactly but gives a difference for D of 48223 Hz, too large to be explained by uncertainties in input data.

Using four data points $T_{1S}(=0)$, $T_{2S}$, $T_{4S}$ and $T_{8S}$, gives a small third term in $1/n^3$ for $\Delta R_{nS}$, absent in (14). The resulting, slightly distorted higher order parabola for $R_{nS}$ gives terms, obeying

$$T'_{nS}=-0{,}0038146973/n^5+4{,}3727722168/n^4-5{,}5572814941/n^3-109677{,}583427429/n^2$$
$$-0{,}0000383854/n+109678{,}771742861 \text{ cm}^{-1} \qquad (17)$$

These reproduce observed data exactly too. The perturbation by a small term in $1/n^5$ does not disprove the essential information on left-right or chiral H behavior (see Section VI).

*IV.2 Case 2: $\sigma \neq 0$*

The uncertainty $\sigma$ for 2S is ±47 Hz but ±10000 Hz for 4S and ±8600 Hz for 8S, about 200 times larger than for 2S (see Table 1). Neglecting small $\sigma$ for 2S, 4S and 8S uncertainties

| 4S/8S | $\|-\sigma(4S)$ | $0\sigma(4S)$ | $+\sigma(4S)$ |
|---|---|---|---|
| $-\sigma(8S)$ | $\|-10000/-8600$ | $0/-8600$ | $+10000/-8600$ |
| $0\sigma(8S)$ | $\|-10000/0$ | $0/0$ | $+10000/0$ |
| $+\sigma(8S)$ | $\|-10000/+8600$ | $0/+8600$ | $+10000/+8600$ |

lead to 9 maximum uncertainty scenarios. Plotting $T_{nS}$ (15) versus terms with uncertainties, redirected to 1S-2S, gives ±10 kHz for 4S/8S $\sigma$-distributions +/+;-/- but these reduce to 50 % for 0/+;0/- and +/0;-/0 and to 10% for +/-.-/+, as shown in Fig. 3. Rounding uncertainties are about ±100 Hz. Numerical results for intermediate uncertainty scenario 0/±, are in Table 3.

**V. Results and Discussion**

Table 2 gives the differences with observed terms in Table 1 for 5 methods A-F, ordered by decreasing average difference for observed terms (I). Results with perturbed parabola (17) and with Kelly data [17] are discussed below. Differences are given in Hz, mainly to illustrate the analytical power in reach. Old and recent QED methods A and B compare well. DWP/QED method C and independent relativistic SDE quantum defect method D are close to DWP theory E of Section III. Columns B and D show the improvement in precision due to DWP fitting. For reproducing observed terms, SDE/DWP methods C-F are superior to QED methods A-B. Table 3 gives $T_{nS}$ with B, D and E for n up to 20. Fig. 4 illustrates $T_{nS}$-differences between method E and all 4 methods A-D. This shows that C, D and E differ in the same way from B, similar to A, see Table 2. $T_{nS}$-differences are within precision limits of Appendix A (within NIST Rydberg uncertainties $\pm 22000(1-1/n^2)$ Hz [20], see Fig. A1. $T_{nS}$ uncertainties with method E, also shown in Fig. 4, are generated by observed uncertainties, following Section IV.2. To be complete, we included $T_{nS}$ differences between E and method F using the perturbed parabola in (17). Digits for $T_{nS}$ in Table 3 merely indicate the accuracy attainable with DWP fitting, as used in E. If a better precision than in Table 1 had been available, this would be reflected immediately in $T_{nS}$ by the analysis of Section III. With 2 nS input terms uncertain by ~10 kHz, discussing differences for $T_{nS}$ would be academic, if it were not for a maximum discrepancy of over 10 kHz [18] for 1S-3S between C-E and B [16], as clearly visible in Fig. 4. The outcome of [19] for will therefore be important for H theories, especially for chiral H theory.

**VI. Hund-type DWP or Mexican hat curve for chiral atom H**

All methods above give a Mexican hat curve for H: curves are either exactly like (1) or good first order approximations thereof, even with QED, based on SDE.

*VI.1 Boltzmann-Hund H Mexican hat curve, based on precise H nS-terms*

The H DWP for all σ=0 derives from equation (15). Level energies $e_{nS}=E_{nS}/c$

$$-e_{nS}=109678{,}771743712/n^2-1{,}188083721/n^2+5{,}555782137/n^3-4{,}36869338/n^4 \text{ cm}^{-1} \quad (18)$$

show large achiral, too symmetric Bohr contribution $\alpha^2/n^2$ or $R_H/n^2$. Bohr could not differentiate between left- or right-handed hydrogen, Coulomb attraction being equal in H ($e^-p^+$) and $\underline{H}$ ($e^+p^-$). The small chiral, less symmetric part, with 3 terms $\alpha^4/n^2$, $\alpha^4/n^3$ and $\alpha^4/n^4$, is responsible for the H fine structure. This is in line with the idea that small left-right differences are only visible when zooming in on a chiral system. The analytical form of the DWP in (18) depends on how term $(R_{\infty S}-R_{1S})/n^2 \approx 1{,}188\ldots/n^2 \text{ cm}^{-1}$ is treated, which refers to the 3 Rydbergs, shown in Fig. 2.

(a) Subtracting $-E_{1S}/n^2$ from (18) gives a first asymmetric H quartic $Q_a$, equal to

$$Q_a=-e_{nS}-t_\infty/n^2=-1{,}188083721/n^2+5{,}555782137/n^3-4{,}36869338/n^4 \text{ cm}^{-1} \quad (19)$$

Since $-E_{1S}$ is used as reference, DWP (19) is the most natural, observed quartic for atom H.

(b) Subtracting $R_{\infty H}/n^2=109677{,}583659991/n^2 \text{ cm}^{-1}$ gives a 2nd asymmetric $Q_b$, equal to

$$Q_b=-e_{nS}-R_{\infty H}/n^2=Q_a+1{,}188\ldots/n^2=+5{,}555782137/n^3-4{,}36869338/n^4 \quad (20)$$

The two asymmetric DWP (a) and (b) only differ by term $1{,}188\ldots/n^2$, following (18)-(20).

(c) Subtracting achiral part $R_{harm}/n^2$ leads to a 3d but symmetric DWP. Coefficients in (18) give

$$b^2=4{,}36869338 \text{ cm}^{-1} \quad (b=\pm 2{,}090141952) \quad (21)$$

$2ab=5,555782137$ cm$^{-1}$, giving $a/b=ab/b^2=0,635863…$ or $n_{c(S)}=b/a=1,572665476$ (22)

$a^2=(½5,555782137/\sqrt{(4,368693379)})^2=1,766358523$ cm$^{-1}$ ($a=±1,329044…$) (23)

Adding a zero term, i.e. by adding and subtracting term $a^2/n^2$ (a particle-antiparticle term of order $\mu\alpha^4$ or $\alpha^2 R_H$) does not alter energy spectrum (2): it gives equivalent $e_{nS}$ and symmetric $Q_c$, obeying

$-e_{nS}=(R_{harm}/n^2)[1-(1,766358523/R_{harm})(1-1,572665476/n)^2]$

$=109679,350018514/n^2-1,766358523(1-1,572665476/n)^2/n^2$ cm$^{-1}$

$Q_c=-e_{nS}-R_{Harm}/n^2=-1,766358523(1-1,572665476/n)^2/n^2$ cm$^{-1}$ (24)

The 3 H DWPs (19), (20) and (24) are graphically presented in Fig. 5. To get at (1), we see that critical $n_{c(S)}=b/a$ (22) for $(a-b/n)^2$ becomes $2n_{c(S)}=2b/a$ for $(a-b/n)^2/n^2$. A plot of $Q_c$ (24) versus

$x=2n_{c(S)}/n-1=2.1,572665476/n-1=3,145330952/n-1$ (25)

shown in Fig. 6, reveals that this 3$^d$ harmonic, symmetric $Q_c$ transforms in

$Q'_c=0,044636093x^4-0,089272185x^2+0,044636093=0,044636093(1-x^2)^2$ cm$^{-1}$ (26)

equivalent with generic $M_x$ (1) in Fig. 1. Dimensionless (26) returns $M_x$ (1) exactly by virtue of

$M_H=Q_c/0,044636093\equiv(1-x^2)^2$ (27)

Result (27) finally and quantitatively validates our straightforward conclusion on chiral atom H in Section II.1 as well as earlier conclusions [12-13]. Uncertainties in Table 1 affect the shape of $M_H$ but slightly and their effect is invisible on the scale of Fig. 6 (0,05 cm$^{-1}$).

*VI.2 H Mexican hat curves from QED/SDE theories*

In fact, H Mexican hat curves with methods A-E all coalesce to the black curve in Fig. 6. In lower resolution, differences between them remain invisible. Even the H Mexican hat curve (red dashes in Fig. 6), generated from much less precise Kelly data [17] in Appendix D, nearly coincides with that for (26). The different $n_{c(S)}$ and $R_{harm}$ values are collected in Table 4.

**VII. Discussion**

(i) Fig. 6 and Table 4 show that all data and H theories available invariantly probe H chiral. With BH theory, chiral H results are self-explanatory and straightforward. Energies $E_{nS}$ in the 2 wells of chiral H reveal that n-values between ∞ and $2n_{c(S)}$ apply for left-handed H-states (Bohr states) with bound e$^-$p$^+$ pair at longer range, whereas those between $2n_{c(S)}$ and 1 gives right-handed <u>H</u>-states for bound e$^+$p$^-$ pair at shorter range (or vice versa). The H e$^-$p$^+$ pair is confined to long range (macroscopic level, H dissociation), in line with Bohr theory. The bound short-range e$^+$p$^-$ pair in <u>H</u> contradicts the long-range process in [1-2], see [22]. It is impossible to explain this H phase transition by n, ℓ, spin ±½, since these remain constant throughout series nS, nP….

(ii) Chiral atom H does not contradict SDE, since this is nearly degenerate with BH closed form theory (see Appendix A). BH theory leads to conceptual and computational advantages for interpreting H fine structure, including a simple approach to the barrier height in the H DWP.

(iii) H 1S-3S, not used as input, should be available with a precision better than 1 kHz [19] to test of chiral H. It would be a better 4[th] reference point but only if less uncertain than term D, used in Section III. This also calls for a better precision for term D itself. The ideal case would be if much more $T_{nS}$ terms were available as precisely as possible ($T_{3S}$-data in Table 3 are in bold).

(iv) A test of whether or not $T_{1S}=0$ is also needed, reminding problems on redirecting $E_{1S}$ [21], as shown also in Table 3. If $T_{1S}\neq 0$, as it seems here, calibration problems may emerge. With the few precise terms now available, $T_{1S}=0$ would require higher order contributions as in (17). Although important, calibration-issues are beyond the scope of this work.

(v) The magnitude of the H fine structure is connected with the coefficient of term $1/n^3$. For nP, this gives $\alpha^2 R_{\infty H} \approx 109677,583…/137,036^2 = 5,84…$ but reduces to 5,55…. for nS. Their ratio gives parameter b (B8) in Appendix B for the Sommerfeld quantum defect.

(vi) H DWPs contain potential difference $(1-n_c/n)^2-1=-2n_c/n+n_c^2/n^2$, i.e. a Kratzer potential [23]. Kratzer was a pupil of Sommerfeld, who already proposed SDE in 1916 [23]. The intimate relation between a classical Mie-type potential like Kratzer's and SDE now easily follows [24].

(vii) Critical $n_{c(S)}$ in (22) and Table 4 is either close to ½π [12,13] or to $2\varphi^{½}= 1,572302756$, where $\varphi=½(\sqrt{5}-1)$ is Euclid's golden number [18]. With $\varphi$, Euclidean H coefficients

$$9\varphi^{½}/4=1,768840600 \text{ and } 2\varphi^{½}=1,572302756 \qquad (31)$$

are close to those in (24). Euclidean H DWP $(9\varphi^{½}/4)(1-2\varphi^{½}/n)^2/n^2$ derives solely from H mass $m_H=m_e+m_P$ [18]. With $\varphi$, concentric circular orbits for nS would also point to spiral behavior (in line with macroscopic cosmological evidence), if not to chaotic behavior [6,18].

(viii) With achiral, polarization independent terms $1/\alpha^2 \approx 2.10^4$ times larger than polarization-dependent chiral terms, the latter will hardly be visible by variations in line intensities. The H spectrum suggests polarization dependent wavelength shifts (PDWSs) [24b].

(viii) For the spectrum of relativistic H (2) to expose generic DWP (1), a zero correction is needed, which seems a contradictio in terminis. In Section II.2, this zero for $E_{nP}$ is qualified as

$$0=+_{1/3}R_H\alpha^2/n^2-_{1/3}R_H\alpha^2/n^2=+(\mu\alpha^4c^2/n^2)6-(\mu\alpha^4c^2/n^2)6=+1,9468-1,9468 \text{ cm}^{-1}- \qquad (32)$$

This zero correction in $1/n^2$ may vanish in SDE (2) but it is essential to get at H Mexican hat curve (1), see Section VI. Knowing symmetry is important for physics [25], this zero term plays an important role, especially when looking at the simplicity of BH chiral DWP H theory.

(ix) Deciding on a person's handedness is difficult when at rest, not perturbed…. Handedness only shows upon performing tasks (eating, writing…). Deciding on H handedness is equally difficult for only one state: it shows when perturbed, when in different fields hν. Unlike [1-2], handedness of natural, neutral, composite system H shows through a n-series, see Fig. 1 and 6.

(x) It appears that, willingly or not, master SDE for bound systems (2), if not the Special Theory of Relativity (STR) in (B11), describes natural left-right differences, following 19[th] century work by Boltzmann, Pasteur, Le Bel and many others [6].

## VIII. Conclusion

Chiral atom H [12,13,25] is given away in any resolution by any theory. SDE, proposed long before spin [23,24], reveals a Boltzmann-Hund H DWP for its fine structure, due to chiral H behavior. More data should reveal whether chiral H symmetry is continuous or discrete in the full n-interval. A Mie-type Kratzer potential for atom H also appears in the potential energy curve (PEC) for bond $H_2$ [26]. As argued before [22,27], the precise generic H DWP found here puts doubts on H-synthesis [1-2] and on common views on neutral matter-antimatter pairs.

**Appendix A *Precision of SDE and Boltzmann-Hund DWP***

The uncertainty for H 2S in Table 1 is $1,9.10^{-14}$. Last digit 3 in observed 2466061413187103(47) Hz [7] is rounded to zero in double precision, giving an error of $1,2165.10^{-15}$. Double precision $10^{-16}$ for $f_{nP}$ times $E_0/h=\mu c^2/h=[m_e/(1+m_e/m_P)]c^2/h=1,23491741822076.10^{20}$ Hz [20] gives uncertainties of order $1,2.10^{20}.10^{-16}$, larger than 10 kHz. Problems with double precision for $f_{nP}$ can easily be quantified using the analytically exact transform $f'_{nP}$ in

$$f_{nP}=1/\sqrt{[1+\alpha^2/(n-\gamma)^2]}\equiv (1-\gamma/n)/\sqrt{[1-2\gamma(1-1/n)/n]}=f'_{nP} \qquad (A1)$$

where Sommerfeld's quantum defect γ for nP is

$$\gamma=1-\sqrt{(1-\alpha^2)}\approx \tfrac{1}{2}\alpha^2+\tfrac{1}{8}\alpha^4+\ldots \qquad (A2)$$

with $1/\alpha=137,035999679$ [20]. Whereas $f_{nP}/f'_{nP}\equiv 1$ or $f_{nP}-f'_{nP}\equiv 0$, differences between $f_{nP}-1$ and $f'_{nP}-1$ are of order $10^{-16}$ in double precision. For n=1 to 300, $E_{nP}$ differences of ±13710 Hz in Fig. A1, are in line with $\pm 22000(1-1/n^2)$ Hz for $R_N=3,289841960361(22).10^{15}$ Hz [20], also shown in Fig. A1. Although double precision $10^{-16}$ is reasonable for physics experiments, master SDE is unreliable for precise H data. Since it is uncertain which of $f_{nP}$ and $f'_{nP}$ is the better option, a closed form quartic is preferable. If close to SDE, this avoids unwanted uncertainties. Fitting SDE (2) in 1/n to 4th, 5th and 6th order, reveals that the best fit is 4th order (5th and 6th give much larger discrepancies, not discussed here). As SDE h.o. start with $\alpha^6/n^6$,

$$-E_{nP}=\tfrac{1}{2}\mu\alpha^2c^2(1/n^2+\alpha^2/n^3-\tfrac{3}{4}\alpha^2/n^4)+\text{h.o.}=\tfrac{1}{2}\mu\alpha^2c^2[1/n^2+(1+\tfrac{1}{2}\alpha^2)\alpha^2(1/n^3-\tfrac{3}{4}/n^4)] \qquad (A3)$$

gives a nP quartic, well within SDE fluctuation limits, as also illustrated in Fig. A1. A factor of $(1+\tfrac{1}{2}\alpha^2)$ for terms $1/n^3$ and $1/n^4$ amply suffices. Skipping this correction gives a difference of 1 MHz for n=1 but it is far from certain whether or not this correction is needed. Linear fits in Fig. A1 reveal that (A3) is closer to $f'_{nP}$ than to $f_{nP}$. For terms (7), $f_{nP}$ and $f_{1P}$ lead to

$$T_{nP}=\mu c^2\{1/\sqrt{[1+\alpha^2/(n-\gamma)^2]}-1/\sqrt{[1+\alpha^2/(1-\gamma)^2]}\}=\mu c^2\{1/\sqrt{[1+(\alpha^2/n^2)(1-\gamma/n)^2]}-\sqrt{(1-\alpha^2)}\} \qquad (A4)$$

(for $T'_{nP}$, $f'_{nP}$ and $f'_{1P}$ apply). Fluctuations for $T_{nP}$ can be twice as large as for $E_{nP}$. We see that (a) closed form (A3) describes in a continuous way (i.e. without fluctuations) the same quantum systems, believed to obey original master SDE (2), the spectrum of the H Dirac Hamiltonian; (b) for precise data, problems occur using fitting with $f_{nP}$ or $f'_{nP}$ with their inverse square root and (c) an accurate low order fit with SDE variable $B^2_{nP}=(\alpha^2/n^2)(1-\gamma/n)^2$ would mean that adding terms to the SDE expansion as in bound state QED [3] could be superfluous.

**Appendix B *Observed H 1S-nS and Sommerfeld's quantum defect method***

We elaborate on (b)-(c) in A. Using Sommerfeld's quantum defect $\gamma_P$, $E_{nP}$ leads to

$$-E_{nP}=\mu c^2\{1/\sqrt{[1+(\alpha^2/n^2)/(1-\gamma/n)^2]}-1\}=\mu c^2[1/\sqrt{(1+B^2_{nP})}-1] \qquad (B1)$$

$$B^2_{nP}=(\alpha^2/n^2)/(1-\gamma/n)^2 \qquad (B2)$$

It is impossible to fit observed terms accurately with $[1/\sqrt{(1+B^2_{nP})}-1]$: its physically meaningless fluctuations are larger than experimental uncertainties, see point (b) above.

An expansion of (B1) in variable (B2) can be truncated to give

$$E_{nP}=\mu c^2(1-\tfrac{1}{2}B^2_{nP}+\tfrac{3}{8}B^4_{nP}-\ldots -1) \tag{B3}$$

Without loss of accuracy, replacing unit 1 by 1/6 leads to a perfect parabola in $B^2_{nP}$, i.e.

$$E_{nP}=\mu c^2(1/6-\tfrac{1}{2}B^2_{nP}+\tfrac{3}{8}B^4_{nP}-\ldots -1/6)=\mu c^2[(\sqrt{6}^{-1}-\sqrt{\tfrac{3}{8}}.B^2_{nP})^2\ldots-6^{-1}] \tag{B4}$$

To apply this for $T_{nS}$, variable (B2) must be parameterized. Without altering α, p would lead to $\gamma_p$ and variable $B_{nS}$, as defined in $\gamma_p=p-\sqrt{(p^2-\alpha^2)}$. As p cancels in first order, we suffice by using b in

$$\gamma_p=1-\sqrt{(1-b\alpha^2)}; \quad B^2_{nS}=(\alpha^2/n^2)/(1-\gamma_p/n)^2 \tag{B5}$$

With b close to 1, differences between nP and nS are quantified (Lamb shifts). $E_{nS}$ and $T_{nS}$ obey

$$E_{nS}=\mu c^2[(\sqrt{6}^{-1}-\sqrt{\tfrac{3}{8}}.B^2_{nS})^2\ldots-6^{-1}]$$

$$T_{nS}=\mu c^2[-\tfrac{1}{2}(B^2_{nS}-B^2_{1S})+\tfrac{3}{8}(B^4_{nS}-B^4_{1S})]=\mu c^2\{-\tfrac{1}{2}(B^2_{nS}-B^2_{1S})[1-\tfrac{3}{4}(B^2_{nS}+B^2_{1S})]\} \tag{B6}$$

($T_{1S}=1/\sqrt{(1+B^2_{nS})}-1/\sqrt{(1+B^2_{1S})}=0$). To fit observed H 1S-nS terms in cm$^{-1}$ with (B6) and to avoid at the same time any the unwanted fluctuations of SDE, we use reduced squared variable

$$B'^2_{nS}=B^2_{nS}/\alpha^2=(1/n^2)/(1-\gamma_p/n)^2 \tag{B7}$$

Since only 3 nS states are precisely known (σ=0), a 2$^{nd}$ order fit in (B7) is exact. This independent relativistic SDE alternative avoids the procedure with running Rydbergs (10) as well as 4$^{th}$ order fitting in 1/n for which 5 data points are needed. As in Section III, we redirect the 2$^{nd}$ order fit for 3 observed nS terms to give exactly term D in Table 1 for non-input data on 3S and 6S. With

$$b=0,951213722\ldots \tag{B8}$$

$-E_{1S}=109678,771743809$ cm$^{-1}$, closer to (15) than that in [16]. For (B7) and 1/n, we get

$$T_{nS}= 4,3684295117855 B'^4_{nS} - 109677,583680585 B'^2_{nS} + 109678,771743809 \text{ cm}^{-1}$$
$$= 4,3656964302/n^4 - 5,5531952381/n^3 - 109677,5842828750/n^2 + 0,0000380725/n +$$
$$109678,7717438090 \text{ cm}^{-1} \tag{B9}$$

SDE terms (B9) for n up to 20 comply with observation (see Tables 2-3, Section IV). Also

$$n_{c(S)}= 1,572318726 \quad \text{and} \quad R_{harm}=109679,3502083 \text{ cm}^{-1} \tag{B10}$$

are close to the values in Section III, especially to Euclidean (32). Even with SDE, 3 terms in $1/n^2$, $1/n^3$ and $1/n^4$ suffice for $T_{nS}$, conforming to DWP results (8)-(10). Assessing term uncertainties for (B9) as in Section IV.2 and Fig. 3, gives similar results as illustrated in Fig. B2. Without quantum defect γ, master SDE (2) or (B1) simplifies to its original STR (Special Theory of Relativity) version [24b]

$$-E_{nP}(STR)/\mu c^2=1/\sqrt{(1+\alpha^2/n^2)}-1= -\tfrac{1}{2}\alpha^2/n^2 +\tfrac{3}{8}\alpha^4/n^4 -\ldots \tag{B11}$$

Since (B11) only adds repulsive term $+\tfrac{3}{8}\alpha^4/n^4$ to Bohr's formula [24b], STR is disproved by the H spectrum. For bound systems (atoms, molecules) like prototypical H, STR does not apply.

## Appendix C *Converting Erickson $E_{nS}$ to $T_{nS}$, with and without Hund fitting*

Old QED $E_{nS}$ data [17] derive from Rydberg $R_E=109737,3177\pm 0,008$ cm$^{-1}$, whereas NIST gives $R_N=109737,31568527$ cm$^{-1}$ [20], meaning $R_N/R_E=1-1,82.10^{-8}$. A direct conversion is obtained from a plot of 1978 $E_{nS}$ versus the 3 precise $T_{nS}$ in Table 1, which gives Erickson $T_{nS}$ set as

$$T_{nS}(conv) = -0{,}999999982080982(-E_{nS}) + 109678{,}771743228 \text{ cm}^{-1} \qquad (C1)$$

This gives differences of order kHz with observed terms. The set is in line with $R_N/R_E$ but is quite different from Reader's [21] (details are not given). Its accuracy compares with recent QED [16], see Table 2, columns A and B. This is not surprising as both use QED and their Rydberg-difference is remedied by (C1).

(i) Adapting original Erickson $E_{nS}$ for a closed form H quartic

$E_{nS}$ do not comply with Boltzmann's DWP or SDE, since a plot versus $1/n$ in 4th order gives

$$-E_{nS} = -4{,}3683378696442/n^4 + 5{,}5554144382477/n^3 + 109677{,}585748732/n^2 - 0{,}000015411526/n \text{ cm}^{-1} \quad (C2)$$

This exposes too large a non-zero term in $1/n$ absent in both theories. A new $E_{1S}$-value gives an equation like (C2) without term $1/n$. A small shift of $6 \cdot 10^{-7}$ cm$^{-1}$ in all $E_{nS}$ levels gives respectively

$$-E_{nS}(DWP) = -109678{,}7737046\ldots - (109678{,}773704 - |E_{nS}|) = 5{,}994 \cdot 10^{-7} - E_{nS} =$$
$$= -4{,}368674993515/n^4 + 5{,}555767059326/n^3 + 109677{,}585628271/n^2 \text{ cm}^{-1} \qquad (C3)$$

This suffices to make Erickson's $E_{nS}$ conform to DWP theory, as remarked previously [18]. This shift of 18 kHz to $E_{nS}$ gives $E_{nS}(DWP)$ but, for $T_{nS}$, this shift cancels throughout.

(ii) Adapting Hund fitted $E_{nS}$ (HU) in (C3) for the Rydberg

As in (C1), a conversion of (C3) provides respectively with

$$T_{nS}(conv,DWP) = -0{,}999999982081926 |-E_{nS}(DWP)| + 109678{,}771743736 \text{ cm}^{-1} \qquad (C4)$$
$$= 109678{,}771743736 - (-4{,}368674993515/n^4 + 5{,}555767059326/n^3$$
$$+ 109677{,}585628271/n^2) \text{ cm}^{-1} \qquad (C5)$$

similar to Boltzmann-Hund DWP (15) and SDE (B8). Results are given in Tables 2-4.

**Appendix D *H DWP from older observed but less precise $T_{nS}$ [17]***

Converting older observed Kelly $T_{nS}$ [17] with the newer more precise terms gives

$$T_{nS}(conv) = 0{,}999999981273885 T_{nS(Kelly)} + 0{,}000040821498226 \text{ cm}^{-1} \qquad (D1)$$

Kelly H DWP parameters are in Table 4 (bottom row). Although Kelly $T_{nS}$ are accurate to 0,0001 cm$^{-1}$ or 3 MHz, they give almost the same H Mexican hat curve, as shown in Fig. 6 (red dashes).

Fig. 1 Generic Boltzmann-Hund DWP or Mexican hat curve for a chiral (quantum) system

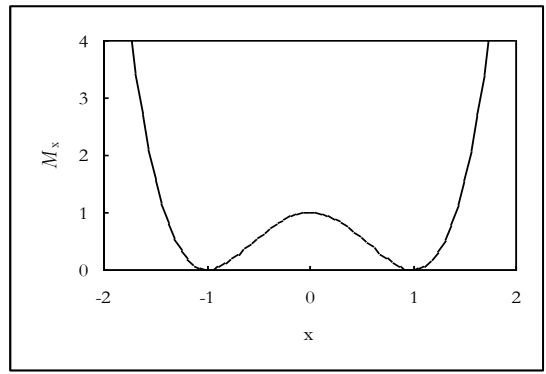

Fig. 2 Parabola for running Rydbergs $R_{nS}$ (in cm$^{-1}$) from 3 observed terms with the 3 asymptotes $R_{\infty S}$, $R_{1S}$ and $R_{harm}$ (lines from top to bottom)

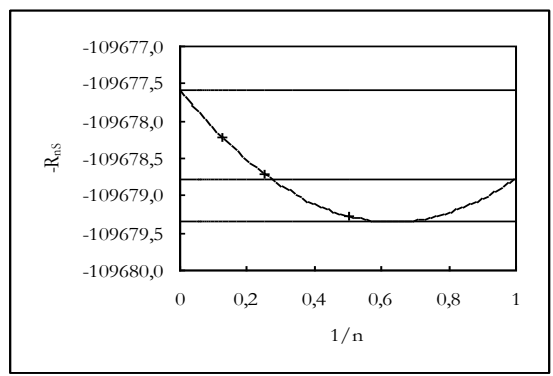

Fig. 3 $T_{nS}$ uncertainties (in Hz) from those reported for 4S and 8S using 2S as reference (top to bottom -/-; 0/-; -/0; +/-; -/+; +/0; 0/+; +/+ as defined in Section IV.2)

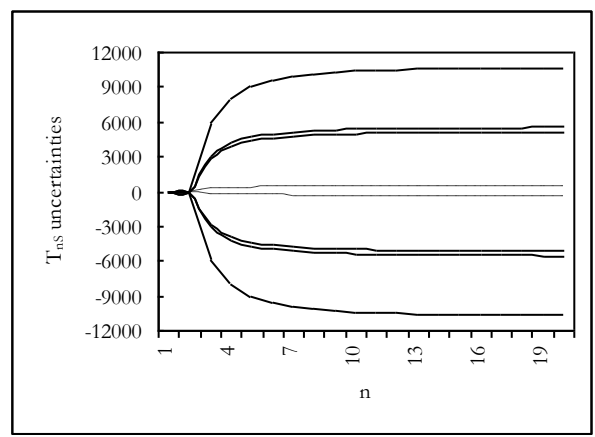

Fig. 4 $T_{nS}$ differences between E and other methods (in Hz) (black: full E-D, dashes E-C; red full E-B, dashes E-A; green E-F)

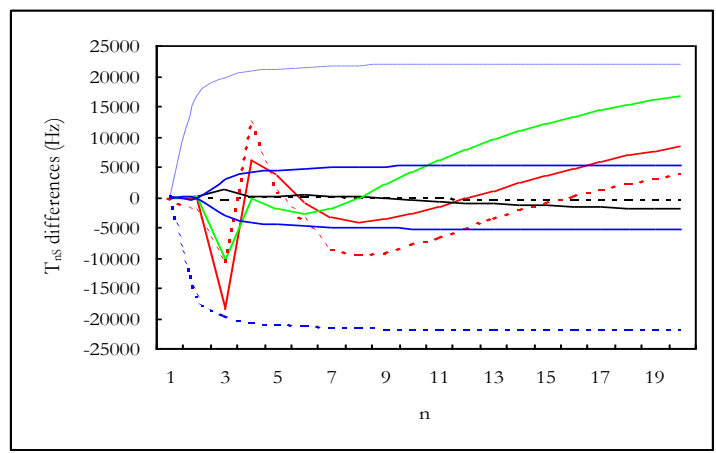

Fig. 5 The 3 H DWPs $Q_a$ (black, natural or Bohr DWP), $Q_b$ (red) and $Q_c$ (blue, Mexican hat curve) (cm$^{-1}$)

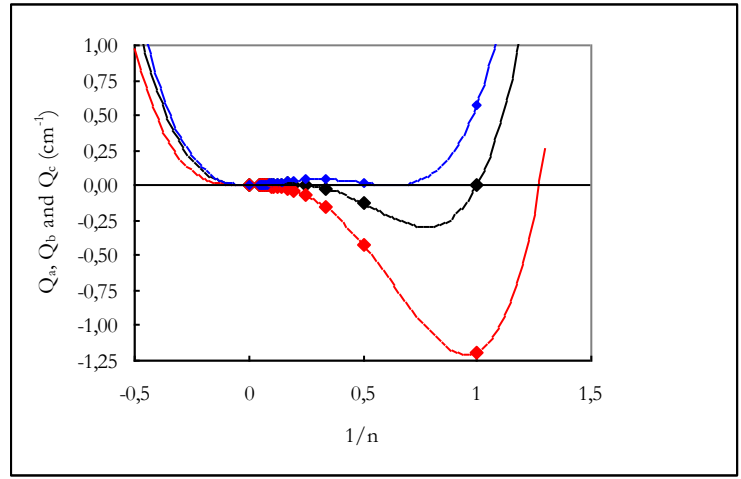

Fig. 6 Generic Mexican hat curve $M_H$ or $Q_c$ using achiral term $R_{harm}/n^2$ (cm$^{-1}$)
(black: this work, red: using Kelly data [17])

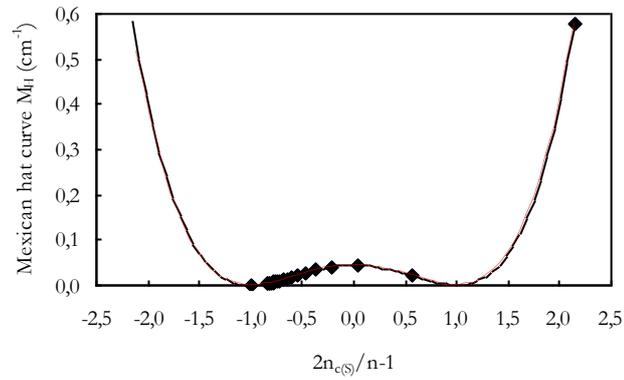

Fig. A1 SDE fluctuations (with linear fits) for n up to 300: ($f_{nP}$-$f'_{nP}$): black (short dashes);
($f_{nP}$-DWP): red (long dashes), ($f'_{nP}$-DWP):green (full line)
NIST Rydberg term uncertainties (blue)

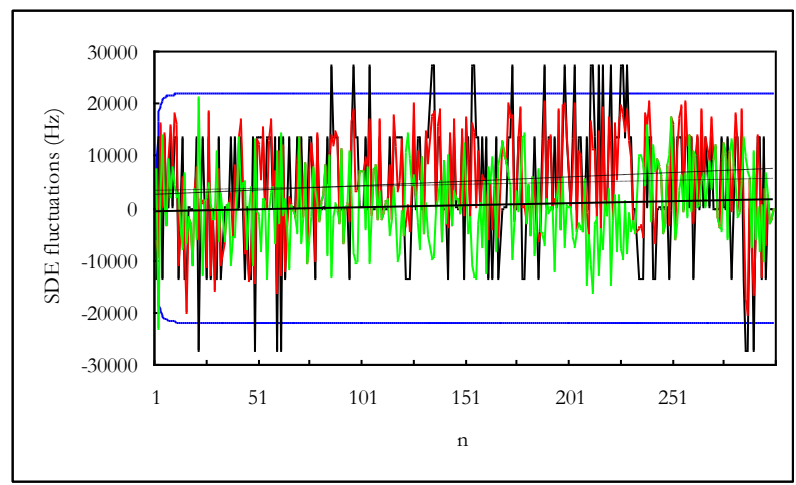

Fig. A2. $T_{nS}$ uncertainties with quantum defect method D with uncertaintiy scenarios in Section IV.2
[red: + + full, - - dashes; green: + - full, - + dashes; blue: + 0 full, 0 + dashes; black:
differences with B (dashes) and E (full); Rydberg uncertainties (outer dashed curves)]

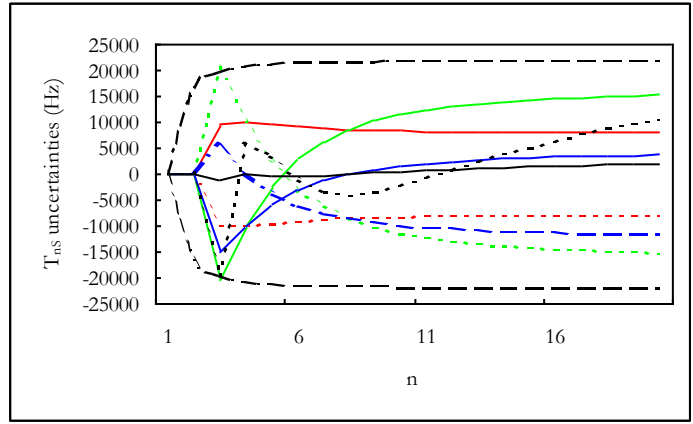

**Table 1.** Observed H 1S-nS, prediction for 1S-3S and differences d with theory (in kHz)

| I. Observed[a] | Value | uncertainty σ | d [16] | d [18] |
|---|---|---|---|---|
| A. 1S-2S [5,7,8] | 2466061413187,103(46) | $1,9.10^{-14}$ | 0,00 | 0,00 |
| B[b]. 2S-8S [9] | 770649350012,0(8,6) | $1,1.10^{-11}$ | -3,98 | 0,00 |
| C. 2S-4S-¼(1S-2S) [10] | 4797338(10) | $2,1.10^{-6}$ | 6,23 | -20,33 |
| D. 2S-6S-¼(1S-3S) [11] | 4197604(21) | $4,9.10^{-6}$ | 3,68 | -6,44 |
| | average d | | (3,7) | (6,7) |
| II. Derived[c] | | | | |
| a. [1S-8S] | [3236710763199,103(*8,6*)] | $2,7.10^{-12}$ | | |
| b[d]. [6S-4S-¼(3S-2S)] | [599734(*31*)] | $5,2.10^{-5}$ | | |
| c. [1S-4S] | [3082581563821(*10*)] | $3,2.10^{-12}$ | | |
| d. [2S-4S] | [616520150635(*10*)] | $1,6.10^{-11}$ | | |
| e. [4S-8S] | [154129199377(*18,6*)] | $1,2.10^{-10}$ | | |

III To be measured [19]
(1S-3S)    current difference is ≈14 kHz: 2922743278672 kHz [16] versus 2922743278657 [18]

[a] the four observed intervals A-D in I. are used for metrology [20]
[b] observed B and derived b, d and e do not depend on 1S (see text)
[c] a=A+B, b=C-D, c=C+(5/4)A…. Errors (in italics) and uncertainties in II derive from those in I.
[d] this interval, uncertain by 31 kHz, does not depend on 1S (like B) but also on unknown 3S and 6S

**Table 2**. Differences with observed I derived II terms in Table 1 with 5 methods A-F (in Hz)

| Method*/Terms | A (C1) | B[16] | C (C5) | D (B9) | E (15) |
|---|---|---|---|---|---|
| **(I) 1S-2S** | **-2431** | **-3** | **1** | **11** | **0** |
| (II) 1S-4S | 12094 | 6231 | 69 | 17 | 0 |
| (II) 1S-8S | -9696 | -4083 | -34 | 3 | 0 |
| **(I) 2S-8S** | **-7265** | **-4081** | **-35** | **-8** | **0** |
| **(I) 2S-4S-0,25(1S-2S)** | **15133** | **6234** | **68** | **3** | **0** |
| **(I) 2S-6S-0,25(1S-3S)** | **1245** | **3679** | **282** | **-2** | **0** |
| (II) 6S-4S-0,25(3S-2S) | 13888 | 2555 | -214 | 5 | 0 |
| Aver diff for (I) | **6518** | **3499** | **96** | **6** | **0** |

A: conversion of old QED $E_{nS}$ (Appendix C)
B QED $T_{nS}$ [16]
C: converting DWP fitted old QED $E_{nS}$ (Appendix C)
D SDE quantum defect (Appendix B)
E: DWP H theory (15) in Section III

**Table 3.** $T_{nS}$ with uncertainties for n up to 20 using methods B (QED [16]) and D-E, this work

| n | B QED [16] | D: (B9) this work | E DWP (15) this work |
|---|---|---|---|
| 1 | | | |
| 2 | 82 258.954 399 283 2(15) | 82 258.954 399 280 8(28) | 82 258.954 399 283 1(0) |
| 3 | **97 492.221 724 658(46)** | **97 492.221 724 005 8(3270)** | **97 492.221 724 045 6(1000)** |
| 4 | 102 823.853 020 867(68) | 102 823.853 021 075(333) | 102 823.853 021 075(135) |
| 5 | 105 291.630 940 843(79) | 105 291.630 940 957(317) | 105 291.630 940 964(151) |
| 6 | 106 632.149 847 323(86) | 106 632.149 847 282(304) | 106 632.149 847 293(160) |
| 7 | 107 440.439 331 863(91) | 107 440.439 331 743(294) | 107 440.439 331 749(165) |
| 8 | 107 965.049 714 599(93) | 107 965.049 714 463(287) | 107 965.049 714 463(169) |
| 9 | 108 324.720 545 876(95) | 108 324.720 545 761(281) | 108 324.720 545 753(171) |
| 10 | 108 581.990 788 289(97) | 108 581.990 788 215(278) | 108 581.990 788 199(173) |
| 11 | 108 772.341 556 766(98) | 108 772.341 556 741(275) | 108 772.341 556 718(174) |
| 12 | 108 917.118 852 716(99) | 108 917.118 852 744(272) | 108 917.118 852 714(175) |
| 13 | 109 029.789 582 86(10) | 109 029.789 582 934(270) | 109 029.789 582 897(176) |
| 14 | 109 119.190 324 18(10) | 109 119.190 324 303(269) | 109 119.190 324 261(176) |
| 15 | 109 191.314 256 35(10) | 109 191.314 256 518(268) | 109 191.314 256 471(177) |
| 16 | 109 250.342 392 65(10) | 109 250.342 392 860(267) | 109 250.342 392 809(177) |
| 17 | 109 299.263 455 09(10) | 109 299.263 455 344(265) | 109 299.263 455 289(178) |
| 18 | 109 340.259 771 78(10) | 109 340.259 772 069(265) | 109 340.259 772 011(178) |
| 19 | 109 374.954 945 76(10) | 109 374.954 946 078(265) | 109 374.954 946 016(178) |
| 20 | 109 404.577 117 11(10) | 109 404.577 117 457(264) | 109 404.577 117 393(178) |
| ∞(-$E_{1S}$) | 109 678.771 743 07 | 109 678.771 743 809(260) | 109 678.771 743 712(180) |
| [redirection of $E_{1S}$ | | 0,001 109 331 26 | 0,000 994 964 037] |

**Table 4.** Critical data for H Mexican hat curves with methods B-E and Kelly data [17]

| Method | $R_{harm}$(cm$^{-1}$) | $n_{c(S)}$ | based on 4$^{th}$ order in 1/n |
|---|---|---|---|
| B: recent QED | 109679,350059184 | 1,572621509 | from $T_{nS}$ in [16] |
| C: older QED | 109679,351984641 | 1,572663125 | from $E_{nS}$ [17], eqn (C5) |
| D: SDE | 109679,3502083 | 1,572318726 | from eqn (B10) |
| E: DWP | 109679,350018514 | 1,572665476 | from eqn (15) |
| [Kelly | 109679,350325 | 1,572373 | from $T_{nS}$ [17], eqn (D1)] |